\documentclass[12pt,preprint]{aastex}
\usepackage{color}

\shorttitle{SNIa luminosity and metallicity}
\shortauthors{Bravo et al.}

\begin{document}

\title{Metallicity as a source of dispersion in the SNIa bolometric light curve
luminosity-width relationship} 
\author{E. Bravo\altaffilmark{1}, I. Dom\'\i nguez\altaffilmark{2}, C.
Badenes\altaffilmark{3,4}, L. Piersanti\altaffilmark{5}, O. Straniero\altaffilmark{5}
}
\altaffiltext{1}{Dept. F\'\i sica i Enginyeria Nuclear, Univ. Polit\`ecnica de
Catalunya, Carrer Comte d'Urgell 187, 08036 Barcelona, Spain;   
eduardo.bravo@upc.edu}
\altaffiltext{2}{Depto. F\'\i sica Te\'orica y del Cosmos, Univ. Granada, 18071 Granada, Spain; 
inma@ugr.es}
\altaffiltext{3}{Benoziyo Center for Astrophysics, Weizmann Institute of Science, Rehovot 76100,
Israel}
\altaffiltext{4}{School of Physics and Astronomy, Tel-Aviv University, Tel-Aviv 69978, Israel;
carles@wise.tau.ac.il}
\altaffiltext{5}{INAF - Osservatorio Astronomico di Teramo,
via mentore Maggini snc, 64100 Teramo, Italy}

\begin{abstract}
The recognition that the metallicity of Type Ia supernova (SNIa) progenitors might bias their use
for cosmological applications has led to an increasing interest in its role on the shaping of SNIa
light curves. We explore the sensitivity of the synthesized mass of $^{56}$Ni,
$M(^{56}\mathrm{Ni})$, to the progenitor metallicity starting from Pre-Main
Sequence models with masses $M_0=2-7$~M$_\odot$ and metallicities $Z=10^{-5}-0.10$. 
The interplay between convective mixing and carbon burning during the 
simmering phase eventually rises the neutron excess, $\eta$, and leads to a smaller
$^{56}$Ni yield, but does not change substantially the dependence of $M(^{56}\mathrm{Ni})$ on $Z$. 
Uncertain attributes of the WD,
like the central density, have a minor effect on $M(^{56}\mathrm{Ni})$.
Our main results are:
1) a sizeable amount of $^{56}$Ni is synthesized during incomplete Si-burning, which leads to a
stronger dependence of $M(^{56}\mathrm{Ni})$ on $Z$ than obtained by assuming that $^{56}$Ni is
produced in material that burns fully to nuclear statistical equilibrium (NSE);
2) in one-dimensional delayed detonation simulations a composition dependence of the
deflagration-to-detonation transition (DDT) density gives a
non-linear relationship
between $M(^{56}\mathrm{Ni})$ and $Z$, and predicts a luminosity larger than previously
thought at low metallicities (however, the progenitor metallicity alone cannot explain the whole
observational scatter of SNIa luminosities), and
3) an accurate measurement of the slope of the Hubble residuals vs metallicity for a large
enough data set of SNIa might give clues to the physics of deflagration-to-detonation transition in
thermonuclear explosions. 
\end{abstract}


\section{Introduction}

In addition to the mass, metallicity is one of the few progenitor attributes
that can leave an imprint on the observational properties of SNIa by affecting the synthesized mass
of $^{56}$Ni, with important consequences for their use as cosmological standard candles. Up to
now, attempts to measure $Z$ directly from
supernova observations have been scarce and their results uncertain \citep{len00,tau08}.
Measuring $Z$ from the X-ray emission of supernova remnants is
a promising alternative but as yet has been only applied to a single supernova
\citep{bad08a}. An alternative venue is to estimate the supernova metallicity as the mean
$Z$ of its environment \citep{bad09}. \citet{ham00} looked for
galactic age or metal content correlations with SNIa luminosity, but their results were
ambiguous. \citet{ell08} looked for systematic trends of SNIa UV spectra
with
metallicity of the host galaxy, and found that the spectral variations were much larger than
predicted by theoretical models. \citet{coo09}, using data
from the Sloan
Digital Sky Survey and Supernova Survey concluded that prompt SNIa are more luminous in metal-poor
systems. Recently,
\citet[hereafter G08]{gal08} and \citet[hereafter H09]{how09}, using different methodologies to
estimate the metallicity of SNIa hosts, arrived to opposite conclusions with respect to the
dependence of supernova luminosity on $Z$. 

There is a long history of numerical simulations of SNIa aimed at predicting
the impact of metallicity and explosive neutronization on their yields
\cite[e.g.][]{bra92,brach00,tra05,bad08a}. \citet[hereafter DHS01]{dom01} found that the
offset in the calibration of supernova magnitudes vs light curve (LC) widths is not monotonic in
$Z$ and remains smaller than $0.07^\mathrm{m}$ for $Z\leq0.02$. \citet{kas09} concluded that the
width-luminosity relationship depends weakly on the metallicity of the progenitor. From an
analytical point of view, \citet[hereafter TBT03]{tim03} using arguments from
basic nuclear physics predicted a linear relationship between $M(^{56}\mathrm{Ni})$ and $Z$. The
conclusions of TBT03 relied on two main assumptions: first, that most of the $^{56}$Ni is
synthesized in material that burns fully to NSE and, second, that a fiducial SNIa produces
a mass 
$M_\mathrm{Fe}^{\eta_0}\sim0.6$~M$_\odot$ of Fe-group nuclei whose $\eta$ is not modified
during the explosion.
\citet{pb08} and \citet{ch08},
based on the same assumptions as TBT03, extended their analysis taking into account the
neutronization produced during the simmering phase.

In this paper, we show that the first assumption of TBT03 does not hold for most SNIa. Indeed,
for a SNIa that produces $M_\mathrm{Fe}^{\eta_0}\sim0.6$~M$_\odot$ the fraction of $^{56}$Ni
synthesized out of NSE exceeds $\sim30\%$. With respect to the second assumption,
\citet[hereafter M07]{maz07} showed, based on observational results, that the mass of Fe-group
nuclei ejected by SNIa spans the range from 0.4 to 1.1~M$_\odot$. This range cannot be accounted
for by metallicity variation within reasonable values.
Accordingly, our working hypothesis is that the yield of $^{56}$Ni in SNIa is governed by a
primary parameter different from $Z$. In our one-dimensional models the primary parameter is the
DDT density, $\rho_\mathrm{DDT}$, although in nature it may be something else such as the
expansion rate during the deflagration phase. The initial metallicity is a secondary factor
that can give rise to scatter in the value of $M(^{56}\mathrm{Ni})$ either directly ({\sl linear
scenario}), by affecting the chemical composition of the ejecta for a given value of the primary
parameter, or indirectly ({\sl non-linear scenario}), by modifying the primary parameter itself.
The understanding of which one of these two characters is actually being played by $Z$ is of
paramount importance. 

\section{The effect of metallicity on the yield of $^ {56}$Ni}

We explore the sensitivity of $M(^{56}\mathrm{Ni})$ to the progenitor metallicity,
starting from Pre-Main Sequence models of masses, $M_0$, in the range $2-7$~M$_\odot$ and
metallicities, $Z$, from $10^{-5}$ to 0.10, as given in the first column of Table~\ref{tab1}. The
initial mass
fractions of all the isotopes with $A\gtrsim6$ have been fixed in solar
proportion, according to \citet{lod03}; consequently, we adopt for the solar metallicity the value
$Z_\odot =0.014$. Each presupernova model has been evolved from the
Pre-Main Sequence to the Thermal Pulse (TP) AGB phase, in order to determine the mass,
$M_\mathrm{core}$, and chemical structure of the C-O core
left behind. Afterwards, an envelope of the appropriate size to reach the Chandrasekhar mass,
$M_\mathrm{Ch}$, has been added on top of the C-O cores, and these structures have been fed as
initial models to a supernova hydrocode. Finally, the explosive nucleosynthesis has been obtained
with a post-processing nucleosynthetic code.

The hydrostatic evolution has been computed by means of the FRANEC code
\citep{chi98}. With respect to the calculations of DHS01, the code has been updated 
in the input physics. For the purposes of the present paper, the most important 
changes concern the $^{12}\mathrm{C}\left(\alpha,\gamma\right)^{16}\mathrm{O}$ reaction rate, which
is calculated according to \citet{kun02} instead of \citet{cf85},  and the treatment of 
convective mixing during the late part of the core-He burning \citep{str03}.

The presupernova model is a Chandrasekhar mass WD built in hydrostatic equilibrium with
a central density $\rho_\mathrm{c}=3\times10^9$~g~cm$^{-3}$. The composition of the envelope of
mass $M_\mathrm{Ch}-M_\mathrm{core}$ is the same as that of the outermost shell of the C-O core.
Thus, instead of assuming C/O=1, as in
DHS01, we adopt the C/O ratio obtained as a result of He-shell burning during the AGB phase.
The effect of changing $\rho_\mathrm{c}$ and the composition
of the envelope has been tested in several models, as explained later. We leave
aside other eventual complexities of pre-supernova physics like rotation \citep{pie03,yoo04}.

The internal composition of the WD is eventually modified during the simmering phase, 
due to the combined effects of convective mixing, carbon burning and electron captures. 
The first two phenomena affect the carbon abundance within the core,
while the latter leads to an increase of $\eta$.
The average (within the WD) carbon consumption and neutron excess increase during the simmering
phase are $\Delta Y(^{12}\mathrm{C})\approx-1.66\times10^{-3}$~mol~g$^{-1}$ and $\Delta\eta
=-\frac{2}{3}\Delta Y(^{12}\mathrm{C})$ \citep{ch08}. We assume that
convective mixing is limited to the C-O core, which implies that the change in the neutron excess
{\sl within the core} is
$\Delta\eta\approx 1.11\times10^{-3}M_\mathrm{Ch}/M_\mathrm{core}$. 
We have also exploded several models disregarding the simmering
phase, to which we will refer in the following as {\sl stratified models}. 

The supernova hydrodynamics code we have used is the same as in \citet{bad03}. As in DHS01, the
present models are based on the delayed-detonation paradigm \citep{kho91}. 
To address the {\sl linear scenario} we take $\rho_\mathrm{DDT}$  
independent of $Z$. In this case, $\rho_\mathrm{DDT}=3\times10^7$~g~cm$^{-3}$,
although
simulations with $\rho_\mathrm{DDT}$ in the range $1-3\times10^7$~g~cm$^{-3}$ are also reported. 

For the {\sl non-linear scenario} we have adopted the criterion that 
a DDT is induced when the laminar flame thickness, $\delta_\mathrm{lam}$, becomes of the order of
the turbulent Gibson length $l_\mathrm{G}$ \citep{rop07}, with the flame properties (velocity and
width) depending on the abundances of $^{12}$C \citep[eq. 22 in][]{woo07b} and
$^{22}$Ne \citep{cha07}, and hence on $Z$ and $\eta$. 
\citet{tow09} concluded from 2D simulations of SNIa that the metallicity 
does not affect the dynamics of the explosion, and so the turbulence intensity is independent of
$Z$. Thus, for a given turbulent intensity a change in $Z$ can be compensated by a change in
$\rho_\mathrm{DDT}$ in order to recover the condition $\delta_\mathrm{lam}/l_\mathrm{G}\approx1$
\citep[see the discussion in][]{cha07}. In this scenario we have scaled the
transition density as a function of the local chemical composition as follows: 
\begin{equation}
\rho_\mathrm{DDT}\propto X(^{12}\mathrm{C})^{-1.3}\left(1+129\eta\right)^{-0.6}\,.
\label{eq2.5}
\end{equation}
\noindent In order to introduce an $\eta$ dependence in the above expression we have assumed, for
simplicity, that the bulk of neutronized isotopes synthesized during the simmering phase
accelerates the carbon consumption rate the same way $^{22}$Ne does.

\subsection{Presupernova evolution}

The results of the hydrostatic evolution of our presupernova models are shown in
Table~\ref{tab1}. For each $M_0$ and $Z$ we give: $M_\mathrm{core}$, the
central abundance of $^{12}$C and $\eta$ in stratified models,
$X_\mathrm{c}(^{12}\mathrm{C})$ and $\eta_\mathrm{c}$, and the same quantities
in the models accounting for the simmering phase,
$X_\mathrm{sim}(^{12}\mathrm{C})$ and $\eta_\mathrm{sim}$. 
In comparison with DHS01, the present models span a larger range of $Z$, as DHS01 computed models
with $Z\lesssim Z_\odot$. In the common range of $Z$ and $M_0$ the results are comparable,
although the adopted rate of the $^{12}\mathrm{C}\left(\alpha,\gamma\right)^{16}\mathrm{O}$
reaction leads to a slightly larger carbon abundance than in DHS01. The differences in
$M_\mathrm{core}$ between
our models and those of DHS01 are smaller than $0.06$~M$_\odot$. The central carbon to oxygen ratio
and $M_\mathrm{core}$ we obtain, and their dependencies with $Z$ and $M_0$, agree as well with
\citet{ume99}.

\subsection{Mass of $M(^{56}\mathrm{Ni})$ ejected}

\begin{figure}
\plotone{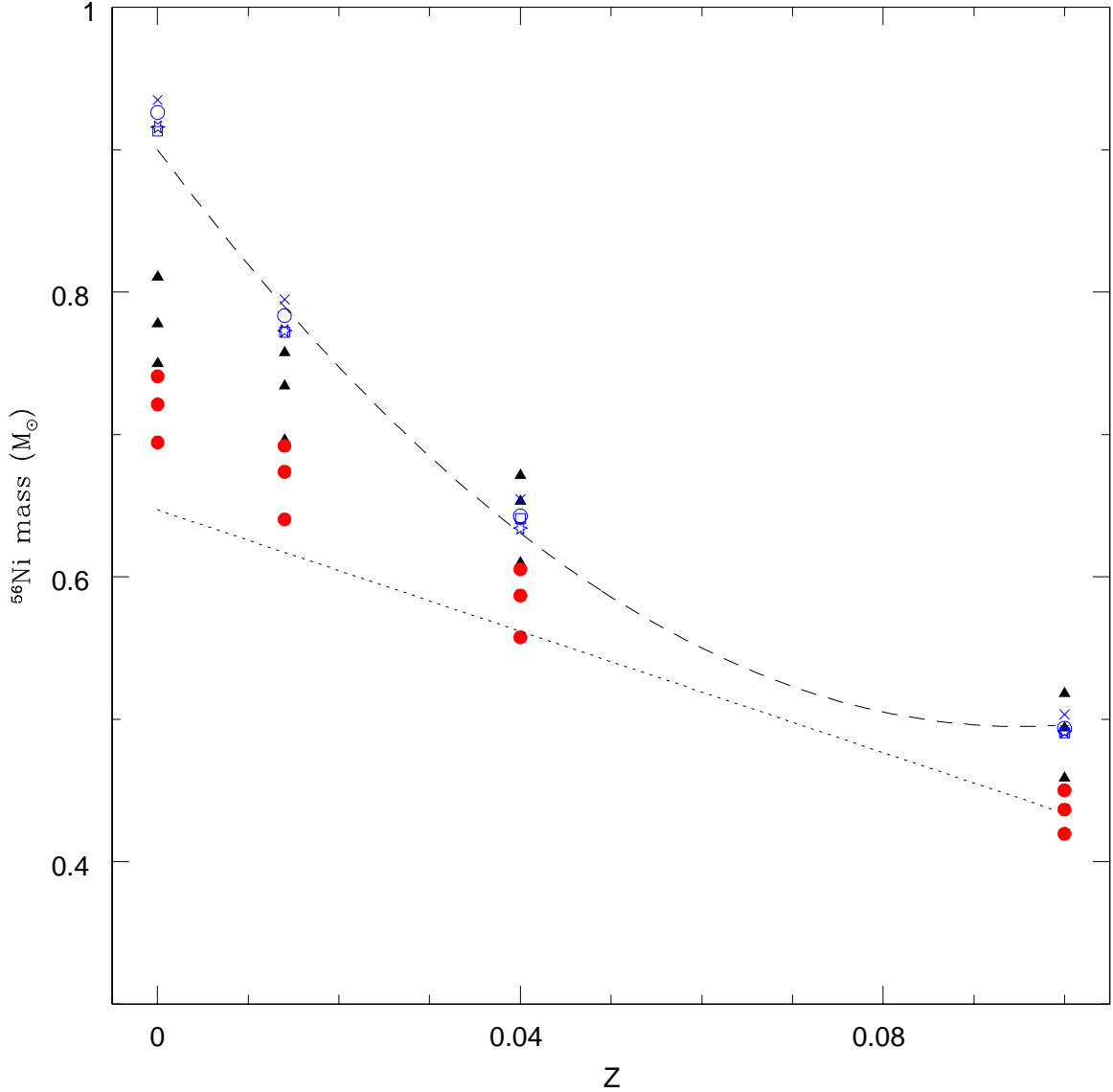}
\caption{
$^{56}$Ni yield vs initial metallicity for different initial masses
and explosion parameters. Black triangles: stratified models with $\rho_\mathrm{c} =
3\times10^9$~g~cm$^{-3}$ and $\rho_\mathrm{DDT} = 3\times10^7$~g~cm$^{-3}$. Red (filled) circles:
models exploded with the same parameters but accounting for the
simmering phase. Blue (empty) circles: the same as red circles except
that $\rho_\mathrm{DDT}$ is a function of the local (at flame position) $Z$ through
$X(^{12}\mathrm{C})$ and $\eta$ (Eq.~\ref{eq2.5}) . Blue crosses and blue stars: the same as
blue circles except that $\rho_\mathrm{c} =
2\times10^9$~g~cm$^{-3}$ (crosses) or $\rho_\mathrm{c} = 4\times10^9$~g~cm$^{-3}$ (stars). Blue
squares: the same as blue circles except that the composition of the
envelope is composed by equal amounts of carbon and oxygen, i.e. C/O = 1 as compared to values
ranging from
C/O = 1.5 to 2.3 as taken from the He-shell burning during the TP phase. The lines
represent the linear relationship between $M
(^{56}\mathrm{Ni})$ and $Z$ proposed in TBT03 (dotted line), and our Eq.~\ref{eq1} (dashed line).
}\label{fig1}
\end{figure}

\begin{figure}
\plotone{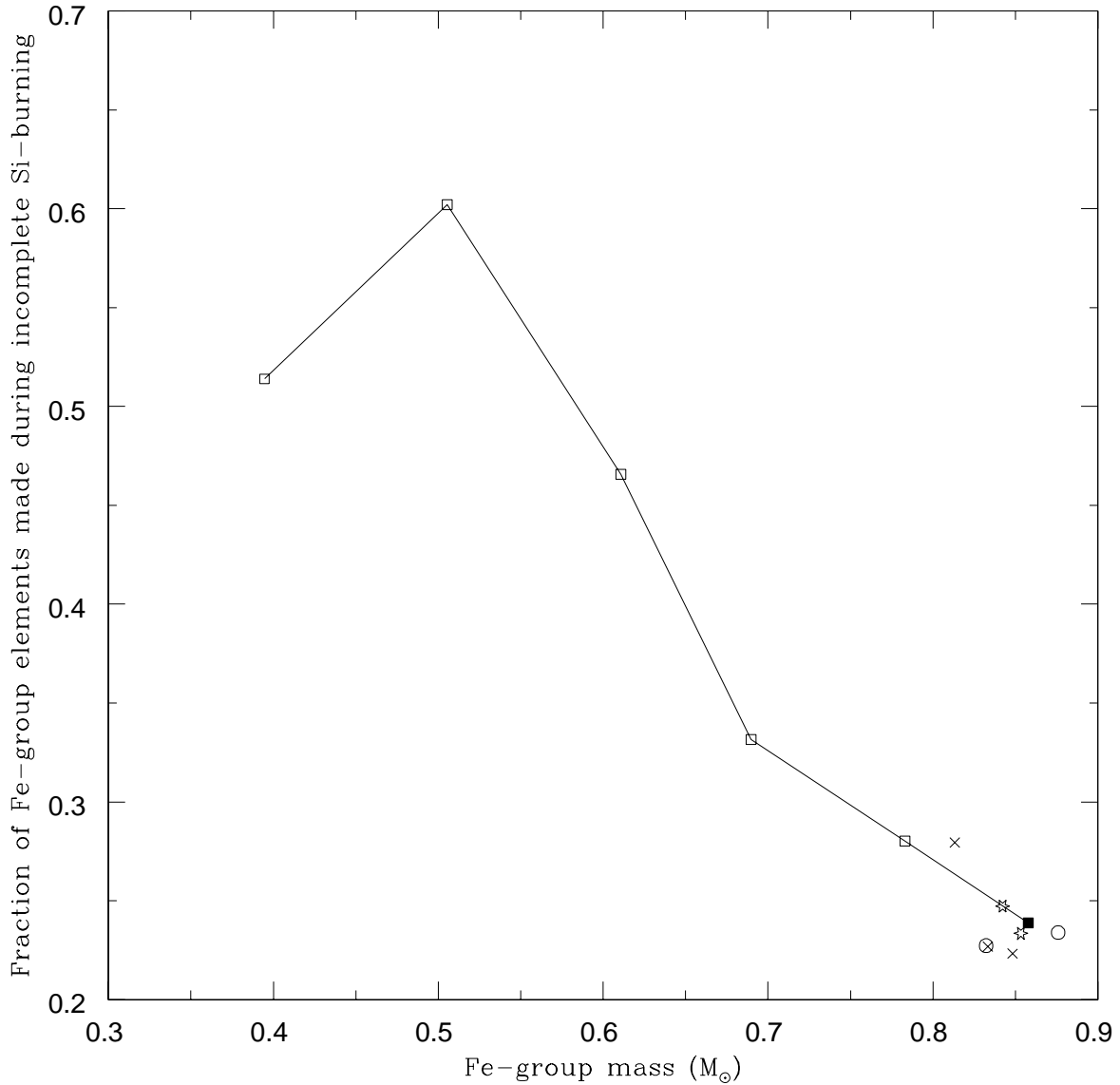}
\caption{
Fraction of Fe-group elements (from Ti to Zn) synthesized in layers experiencing incomplete
Si-burning ($T_\mathrm{max}\leq5.2\times10^9$~K) as a function of the total mass of Fe-group
elements ejected. The filled square is a reference model with $\rho_\mathrm{DDT} =
3\times10^7$~g~cm$^{-3}$, $Z=10^{-5}$, and $M_0 = 5 \mathrm{M}_\odot$, that accounts for the
simmering phase. The empty squares belong to models with varying
$\rho_\mathrm{DDT}$, from $2.6\times10^7$ to $10^7$~g~cm$^{-3}$. All the above models are linked
by a solid line to help guiding the eye. The rest of models show the
sensitivity to different model parameters with respect to the reference model, as follows:
crosses, varying $Z$; empty circles, varying $M_0$; stars, varying $\rho_\mathrm{c}$
}\label{fig2}
\end{figure}

The results of the explosion simulations are summarized in Figs.~\ref{fig1} and
\ref{fig2}. Figure~\ref{fig1} shows the dependence of $M(^{56}\mathrm{Ni})$ on $Z$. For the
stratified models, we obtain the same range of variation of
$M(^{56}\mathrm{Ni})$ with respect to $M_0$ at given $Z$ as DHS01: 0.06~M$_\odot$,
although
the $^{56}\mathrm{Ni}$ yields do not match with DHS01 because they used different values of
$\rho_\mathrm{c}=2\times10^9$~g~cm$^{-3}$ and $\rho_\mathrm{DDT}=2.3\times10^7$~g~cm$^{-3}$.
The models accounting for the simmering phase behave like the stratified models with
respect to variations in
$M_0$ and $Z$, although with a smaller total $M(^{56}\mathrm{Ni})$
due to electron captures during the simmering phase. The dependence of
$M(^{56}\mathrm{Ni})$ on $Z$ can be approximated by a linear function:
\begin{equation}
M(^{56}\mathrm{Ni})\propto f(Z)=1-0.075\frac{Z}{Z_\odot}\,,
\label{eq0}
\end{equation}
\noindent while stratified models can be approximated by:
$M(^{56}\mathrm{Ni})\propto1-0.069Z/Z_\odot$, i.e. the slope of the linear function is quite
insensitive to the carbon simmering phase.

To explore the {\sl non-linear scenario} we
have computed models accounting for the simmering phase with fixed $M_0 = 5$~M$_\odot$. 
Introducing a composition dependent $\rho_\mathrm{DDT}$ produces a qualitatively different result
because the relationship between
$M(^{56}\mathrm{Ni})$ and $Z$ is no longer linear, especially at low metallicites for which a
larger
$\rho_\mathrm{DDT}$ is obtained, implying a much larger $M(^{56}\mathrm{Ni})$. Our results can
be fit by a polinomial law:
\begin{equation}
M(^{56}\mathrm{Ni})\propto f(Z)=1-0.18\frac{Z}{Z_\odot}\left(1-0.10\frac{Z}{Z_\odot}\right)\,.
\label{eq1}
\end{equation}
Both the central density at the onset of thermal runaway and the final C/O ratio in the accreted
layers have a minor effect on $M(^{56}\mathrm{Ni})$ within the explored range.

TBT03 proposed a linear relationship between
$M(^{56}\mathrm{Ni})$ and $Z$: $M(^{56}\mathrm{Ni})\propto1-0.057Z/Z_\odot$ (dotted line in
Fig.~\ref{fig1}). In all of our present models we find a steeper slope. The reason for this
discrepancy lies in the assumption by TBT03 that most of the $^{56}$Ni is synthesized in
NSE. In our models a sizeable fraction of $^{56}$Ni is always synthesized during incomplete
Si-burning, whose final composition has a stronger dependence on $Z$ than NSE matter. 
As \citet{hix96} showed, the mean neutronization of Fe-peak isotopes during
incomplete Si-burning is much larger than the global neutronization of matter because neutron-rich
isotopes within the Si-group are quickly photodissociated, providing free neutrons that are
efficiently captured by nuclei in the Fe-peak group, favouring their neutron-rich isotopes.
Figure~\ref{fig2} shows that up to $\sim60\%$ of $M_\mathrm{Fe}^{\eta_0}$ can be made out
of NSE, the actual fraction depending essentially on the total mass of Fe-group elements ejected.
Thus, the less $M(^{56}\mathrm{Ni})$ is synthesized, the larger fraction of it is built during
incomplete Si-burning and the stronger is its dependence on $Z$.

\section{Discussion}

The results presented in the previous section show that the metallicity is not the primary
parameter that allows to reproduce the whole observational scatter of $M(^{56}\mathrm{Ni})$, for a
reasonable range of $Z$. We have also shown that a possible dependence of the primary parameter on
$Z$, would lead to a non-linear relationship between $M(^{56}\mathrm{Ni})$ and $Z$, as in
Eq.~\ref{eq1}. However, as we will show in the following, it would be possible to unravel the way
$M(^{56}\mathrm{Ni})$ depends on $Z$ by means of future accurate measurements of SNIa properties.

\begin{figure}
\plotone{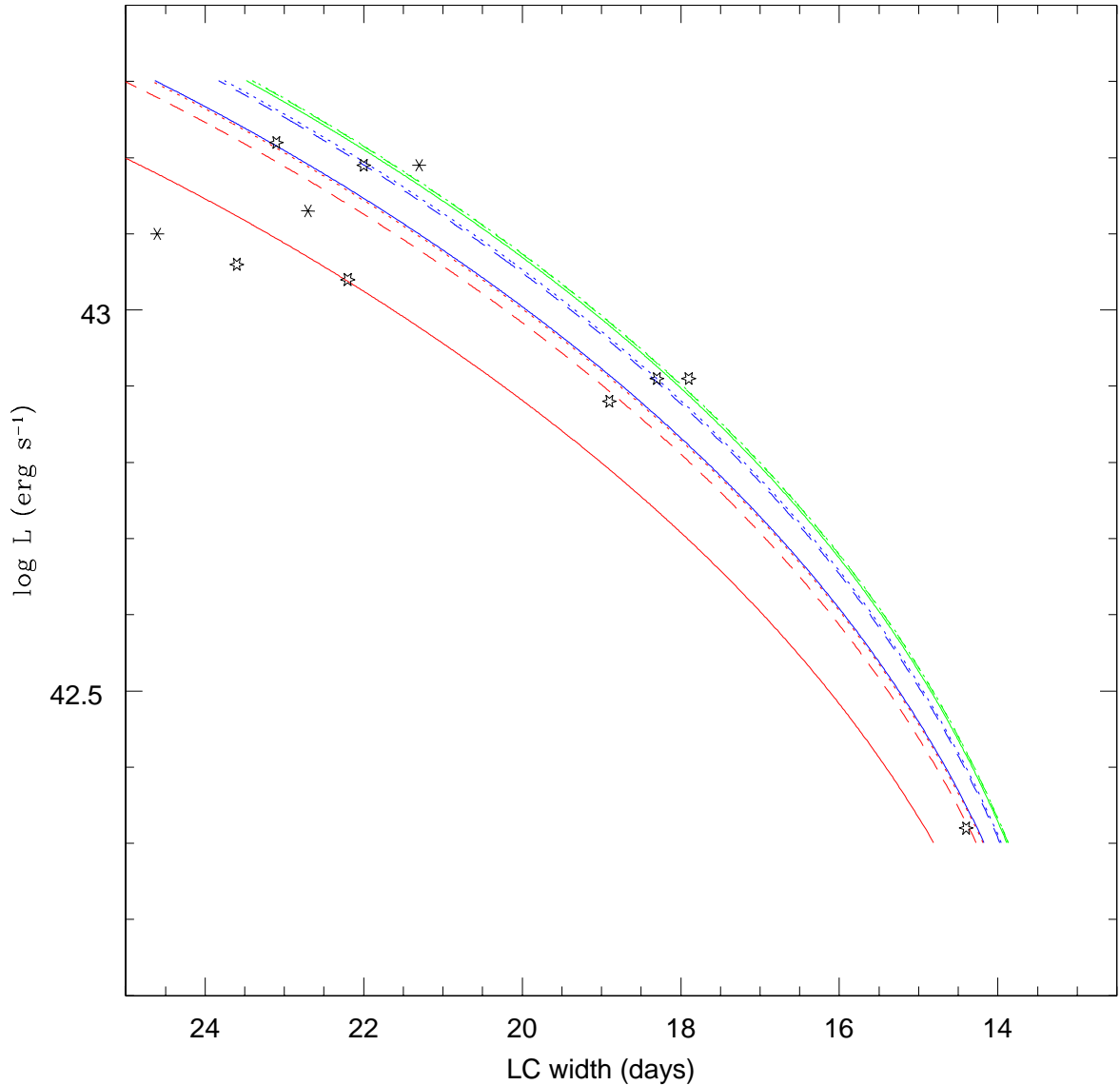}
\caption{
Sensitivity of the bolometric light curve luminosity-width relationship to the initial metallicity.
The solid lines belong to Eq.~\ref{eq1} of the present work for $Z=0.1 Z_\odot$ (green),
$Z=Z_\odot$
(blue), and $Z=3 Z_\odot$ (red). There are also shown the curves obtained using the
$M(^{56}\mathrm{Ni})$ vs Z relationships of TBT03 (dotted lines) and H09 (dashed
lines). The stars represent SNIa from \citet{con00}, whose light-curve width has been
computed as in M07, while the asterisks represent SN1999ac \citep{phi06}, SN2003du \citep{sta07},
and SN2005df \citep{wan09} (by order of decreasing LC width), estimated from the published
bolometric light curves 
}\label{fig3}
\end{figure}

We start analysing the amount of the scatter induced by the dependence of $M(^{56}\mathrm{Ni})$ on
$Z$ given by Eq.~\ref{eq1}. For simplicity we follow the procedure of M07 to
estimate the supernova luminosity and LC width. The peak bolometric
luminosity, $L$, is determined directly by the mass of $^{56}$Ni synthesized (in the following,
all masses are in $M_\odot$ and energies are in $10^{51}$~ergs):
\begin{equation}
L\left[M(^{56}\mathrm{Ni})\right]=2\times10^{43}M(^{56}\mathrm{Ni})~\mathrm{erg}~\mathrm{s}^{-1}\,,
\label{eq1.5}
\end{equation}
\noindent while the bolometric LC width, $\tau$, is determined by the kinetic energy,
$E_\mathrm{k}$, and the opacity, $\kappa$: $\tau\propto\kappa^{1/2}E_\mathrm{k}^{-1/4}$.
The kinetic energy is given by the difference of the WD initial binding energy,
$\mathrm{|BE|}$, and the nuclear energy released, the latter being related to the final chemical
composition of the ejecta:
$E_\mathrm{k}
\approx1.56M(^{56}\mathrm{Ni})+1.74\left[M_\mathrm{Fe}-M(^{56}\mathrm{Ni})\right]+1.24M_\mathrm{IME
}-\mathrm{|BE|}$ , where
$M_\mathrm{Fe}$ is the total mass of Fe-group nuclei and $M_\mathrm{IME}$ is the mass of
intermediate-mass elements (IME). The
opacity is provided mainly by Fe-group nuclei and
IMEs: $\kappa\propto M_\mathrm{Fe}+0.1M_\mathrm{IME}$. We have taken $\mathrm{|BE|}=0.46$, which is
a good approximation given the small variation of binding energy with initial central density:
$\mathrm{|BE|}$ is in the range $0.44-0.47$ for $\rho_\mathrm{c}=2-4\times10^9$~g~cm$^{-3}$. To
reduce the number of free parameters we further link $M_\mathrm{IME}$ to $M_\mathrm{Fe}$ imposing
that
the ejected mass is the Chandrasekhar mass ($M_\mathrm{Ch}\approx1.38~\mathrm{M}_\odot$ in our
models), and that the amount of unburned C+O scales as $M_\mathrm{CO}\approx0.3M_\mathrm{IME}^2$,
as
deduced from our models. Thus,
$M_\mathrm{Fe}+M_\mathrm{IME}+0.3M_\mathrm{IME}^2=M_\mathrm{Ch}$. Furthermore, the mass of
$^{56}$Ni is linked to the mass of Fe-group nuclei by
$M(^{56}\mathrm{Ni})=M_\mathrm{Fe}^{\eta_0}\times
f(Z)=\left(M_\mathrm{Fe}-M_\mathrm{ec}\right)\times
f(Z)$, where $f(Z)$ is given by
Eq.~\ref{eq1} or a similar function, and $M_\mathrm{ec}$ is the mass of the neutron-rich
Fe-group core (due to electron captures during the explosion). We have taken
$M_\mathrm{ec}\approx0.14~\mathrm{M}_\odot$, which is representative of the range of masses
obtained in our models: 
$0.10-0.16~\mathrm{M}_\odot$ for $\rho_\mathrm{c}=2-4\times10^9$~g~cm$^{-3}$. Finally, to compare
with observed values a scale factor of 24.4 is applied to the value of $\tau$ thus obtained, as in
M07.
Putting all these together, we obtain the
following expression for the bolometric LC width (in days) as a function of $M(^{56}\mathrm{Ni})$
and $Z$:
\begin{equation}
\tau\left[M(^{56}\mathrm{Ni}),Z\right] = 21.9\frac
{\left\lbrace\displaystyle{\frac{M(^{56}\mathrm{Ni})}{f(Z)}}-0.027+0.263\sqrt{1-0.482\displaystyle{
\frac{M(^{56}\mathrm{Ni})} {f(Z)}}}\right\rbrace^{1/2}}
{\left\lbrace\left(\displaystyle{\frac{1.115}{f(Z)}}-0.115\right)M(^{56}\mathrm{Ni})-1.46+2.09\sqrt
{ 1-0.482\displaystyle{\frac{M(^{56}\mathrm{Ni})}{f(Z)}}}\right\rbrace^{1/4}}\,.
\label{eq2}
\end{equation}
\noindent The relationship between $L$ and $\tau$ derived from Eqs.~\ref{eq1}, \ref{eq1.5} and
\ref{eq2} is displayed in Fig.~\ref{fig3} for three different metallicities along
with observational data. There are also represented the relationships obtained by substituting
Eq.~\ref{eq1} by the
$M(^{56}\mathrm{Ni})$ vs $Z$ dependences proposed by TBT03 and Eq.~5 in
H09. Our Eq.~\ref{eq1} gives a wider range of
$M(^{56}\mathrm{Ni})$, which accounts better for the scatter of the 
observational data. Indeed, if real SNIa follow Eq.~\ref{eq1}, deriving supernova
luminosities from $Z$-uncorrected LC shapes might lead to systematic errors of up to
0.5~magnitudes.

To estimate the bearing that the metallicity dependence of $M(^{56}\mathrm{Ni})$ can have on
cosmological studies that use a large
observational sample of supernovae, we have generated a
virtual population of 200 SNIa that has been analyzed following the same methodology as
G08 and H09. Each virtual supernova has been randomly
assigned a progenitor
metallicity, from a uniform distribution of $\log(Z)$ between
$Z_\mathrm{min}=0.1Z_\odot$ and $Z_\mathrm{max}=3Z_\odot$, and an 
$M_\mathrm{Fe}$, uniformly distributed in the range from
0.31 to 1.15~$\mathrm{M}_\odot$. The minimum and maximum $M(^{56}\mathrm{Ni})$ thus obtained
(computed with Eq.~\ref{eq1} and $M_\mathrm{ec}=0.14~\mathrm{M}_\odot$) are
$0.1$ and $1~\mathrm{M}_\odot$, and the bolometric LC width, $\tau$, lies in the range
$15-24$~days. A $Z$-uncorrected mass of $^{56}$Ni, $M^{\odot}_{56}$, has then been
obtained as the value of $M(^{56}\mathrm{Ni})$ that would give the same $\tau$ if $Z=Z_\odot$. The
$M^{\odot}_{56}$ so computed gives an idea of the effect of fitting an observed SNIa LC with a
template that takes no account of the supernova metallicity. From Eq.~\ref{eq1.5}, we estimate the
Hubble
Residual, HR, of each virtual
SNIa at: $\mathrm{HR}=2.5\log\left(M^{\odot}_{56}/M(^{56}\mathrm{Ni})\right)$. As a final step
we have added gaussian noise with $\sigma=0.1$ to both HR and $\log(Z)$, to simulate the effect of 
observational uncertainties. 

\begin{figure}
\plotone{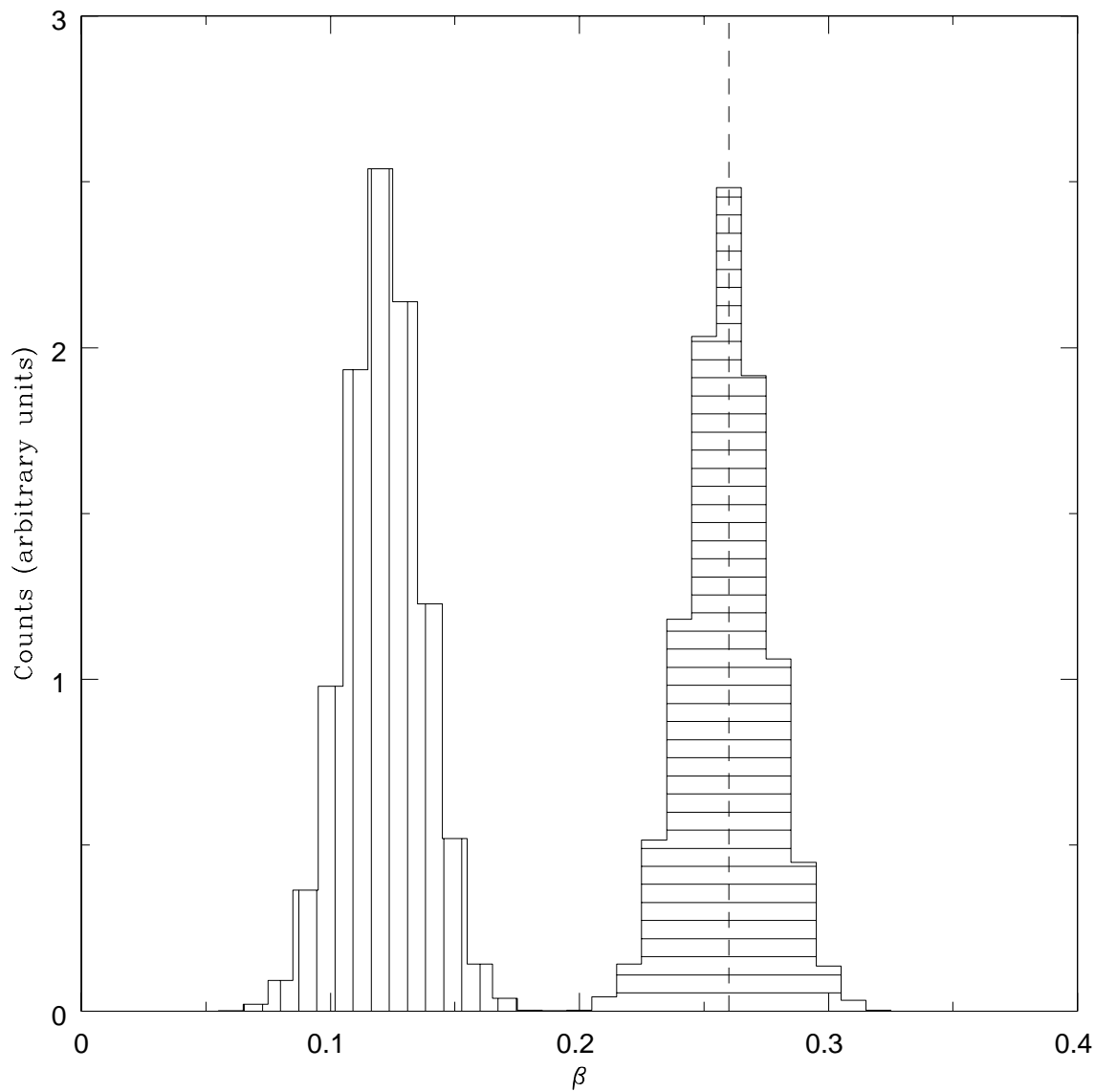}
\caption{
Number distribution of the slope, $\beta$, of the least-squares fit to the Hubble residual vs
metallicity. The statistics used 10,000 data sets, each one obtained adding gaussian noise
($\sigma=0.1$ in HR, and 0.1~dex in $Z$) to a virtual random population of 200 SNIa generated
using either the quadratic $M(^{56}\mathrm{Ni})$ vs Z relationship (Eq.~\ref{eq1}, horizontally
hatched
histogram) or the linear relationship (Eq.~\ref{eq0}, vertically hatched histogram). The vertical
dashed
line shows the slope measured by G08
}\label{fig4}
\end{figure}

A linear relationship $\mathrm{HR}=\alpha+\beta\log(Z)$ has then been fit to the
noisy virtual data by the least-squares technique, as in G08 and H09. Figure~\ref{fig4} shows the
results for
10,000 realizations of the noisy virtual dataset. The histogram gives the number counts of
the slope $\beta$ in the 10,000 realizations. The whole process
has been repeated by using
Eq.~\ref{eq0} (i.e. the {\sl linear scenario}) to represent the dependence of $M(^{56}\mathrm{Ni})$
on $Z$ and the results are also
shown in
Fig.~\ref{fig4}. From the Figure it is clear that, for a large
enough set of SNIa whose luminosity and metallicity are measured with small enough errors, it is
possible to discriminate between the {\sl linear} and {\sl non-linear scenarios}. 
In our numerical experiment, the mean value of $\beta$ is 0.13 in
the first case and 0.26 in the second case, both with a standard deviationof 0.02.

Figure~\ref{fig4} shows also the observational results obtained by G08, who
approximated the metallicities of the SNIa in their sample by the $Z$ of the host
galaxy, obtained from an empirical galactic mass-metallicity relationship. The striking match
between our results based on the {\sl non-linear scenario} and those of G08 must be viewed with
caution in view of the observational uncertainties involved in measuring supernova
metallicities and the limitations of our models (i.e. the assumption of spherical symmetry).
Recently, using a different method of determination of the SNIa metallicity,
H09 arrived to a result opposite to that of G08, i.e. they
found that HR is uncorrelated with $Z$, leading to a distribution centered around
$\beta\approx0$. Thus, until
such discrepancies are resolved it is not possible to draw any firm conclusion about the
metallicity effect on SNIa luminosity. However, it is worth stressing that the simultaneous
measurement of supernova luminosity and metallicity for a large SNIa set would strongly constrain
the physics of the deflagration-to-detonation transition in thermonuclear
supernovae, one of the key standing problems in supernova theory. 

\acknowledgments
This work has been partially supported by the MEC grants AYA2007-66256 and AYA2008-04211-C02-02, by
the European Union FEDER funds, by the Generalitat de Catalunya, and by the ASI-INAF I/016/07/0. CB
thanks Benoziyo Center for Astrophysics for support

\bibliographystyle{aa}

\clearpage
\centering
\begin{deluxetable}{llccccc}
\tabletypesize{\normalsize}
\tablecaption{Properties of CO cores at the beginning of TPs}
\tablecolumns{7}
\tablewidth{0pt}
\tablehead{
\colhead{$Z$\tablenotemark{a}} & \colhead{$M_0$} & \colhead{$M_\mathrm{core}$} & 
\colhead{$X_\mathrm{c}$ ($^{12}$C)} & \colhead{$\eta_\mathrm{c}$} & 
\colhead{$X_\mathrm{sim}$ ($^{12}$C)} & \colhead{$\eta_\mathrm{sim}$}
\\
& ($\mathrm{M}_\odot$) & ($\mathrm{M}_\odot$) & & & & \\
}
\startdata
$10^{-5}$ (0.23) & 3 & 0.801 & 0.28 & $8.7\times10^{-7}$ & 0.38 & $1.9\times10^{-3}$ \\
$10^{-5}$ (0.23) & 5 & 0.903 & 0.25 & $8.6\times10^{-7}$ & 0.31 & $1.7\times10^{-3}$ \\
$10^{-5}$ (0.23) & 6.5 & 1.052 & 0.20 & $8.4\times10^{-7}$ & 0.24 & $1.5\times10^{-3}$ \\
0.014 (0.269) & 3 & 0.615 & 0.23 & $1.2\times10^{-3}$ & 0.36 & $3.7\times10^{-3}$ \\
0.014 (0.269) & 5 & 0.848 & 0.32 & $1.2\times10^{-3}$ & 0.38 & $3.0\times10^{-3}$ \\
0.014 (0.269) & 7 & 1.005 & 0.22 & $1.2\times10^{-3}$ & 0.26 & $2.7\times10^{-3}$ \\
0.040 (0.31) & 3 & 0.608 & 0.18 & $3.4\times10^{-3}$ & 0.33 & $6.0\times10^{-3}$ \\
0.040 (0.31) & 5 & 0.843 & 0.32 & $3.4\times10^{-3}$ & 0.39 & $5.4\times10^{-3}$ \\
0.040 (0.31) & 7 & 1.032 & 0.28 & $3.4\times10^{-3}$ & 0.30 & $5.1\times10^{-3}$ \\
0.10 (0.38) & 3 & 0.624 & 0.15 & $7.3\times10^{-3}$ & 0.29 & $10.1\times10^{-3}$ \\
0.10 (0.38) & 5 & 0.846 & 0.29 & $7.6\times10^{-3}$ & 0.34 & $9.6\times10^{-3}$ \\
0.10 (0.38) & 7 & 0.962 & 0.23 & $7.6\times10^{-3}$ & 0.26 & $9.5\times10^{-3}$ \\
\enddata
\tablenotetext{a}{Initial metallicity (and helium abundance)}
\label{tab1}
\end{deluxetable}

\clearpage

\clearpage

\clearpage

\clearpage

\end{document}